\documentclass[aps,prd,twocolumn,floatfix,nofootinbib,showpacs,superscriptaddress]{revtex4-1}

\usepackage{float,amsmath,amsfonts,amssymb,slashed,bm,graphicx,color}

\begin{document}

\title{Interface tension and interface entropy in 2+1 flavour Nambu-Jona-Lasino model}       

\author{Wei-yao Ke}   
\affiliation{School of Physics and State Key Laboratory of 
Nuclear Physics and Technology, Peking University, Beijing 100871,
China}
\author{Yu-xin Liu}
\affiliation{School of Physics and State Key Laboratory of
Nuclear Physics and Technology, Peking University, Beijing 100871,
China} 
\affiliation{Collaborative Innovation Center of Quantum Matter, Beijing, China} 
\affiliation{Center for High Energy Physics, Peking
University, Beijing 100871, China}

\date{\today}

\begin{abstract}
We study the QCD phases and their transitions in 2+1 flavour NJL model, with focus on the interface effects such as the interface tension, the interface entropy and critical bubble size in the coexistence region of the first order phase transitions. Our results show that, under the thin-wall approximation, interface contribution to total entropy density changes its discontinuity scale in the first order phase transition. However, the entropy density of dynamical chiral symmetry (DCS) phase is always greater than that of the dynamical chiral symmetry broken (DCSB) phase in both heating and hadronization processes. To address this entropy puzzle, the thin-wall approximation is evaluated in the present work. We find that the puzzle can be attributed to a drastic overestimate of the critical bubble size at low temperature in the hadronization process. With an improvement on the thin-wall approximation, the  entropy puzzle is well solved with the total entropy density of the hadron-DCSB phase exceeding apparently that of the DCS-quark phase at low temperature.
\end{abstract}

\maketitle

\section{Introduction}

The phase transitions and phase diagram of QCD have been attracting great attentions from experimental and theoretical physicists in recent years~\cite{Wambach:2009Review,Fukushima:2011Review}. Based on present knowledge, a chiral symmetry restoration phase transition and a deconfinement phase transition can be defined in the vanishing quark mass limit and infinite quark mass limit respectively \cite{Fukushima:2008PNJL}. While at physical mass, the natures of these phase transitions are under exploration. The first principle, lattice QCD calculations indicate that the temperature driven phase transition at vanishing chemical potential is a crossover \cite{Aoki:2006LatticeCrossover}. Although no conclusive results from lattice QCD are obtained at higher chemical potential (of the order of temperature) due to the sign problem, not only many effective model calculations~\cite{Sasaki:2008NJL,Costa:2008NJL,Jiang:2013NJL,Fu:2008PNJL,Ciminale:2008PNJL,Fukushima:2008PNJL,CEP:2008PNJL,Sasaki:2011CEPPNJLMassE,Schaefer:2007PQM,HuangMei:2010sigmaM,Kovacs:2008ChiralQM,Hatta:2003GL} but also the Dyson-Schwinger equation approach of continuum QCD~\cite{Qin:2011DSE,Fischer:2011DSE,Zong:2009DSE}  have shown that there is a chemical potential driven first order phase transition at low temperature and, as the temperature goes higher, the strength of the first order transition decreases and finally becomes a crossover at and beyond the critical end point (CEP).

Interesting behavior of first order phase transition is the phase coexistence. When first order phase transition takes place, the system is inhomogeneous and two phases with different particle numbers, energy and entropy densities meet each other at an interface. The formation of this inhomogeneity is guided by a slow nucleation or an explosive spinodal decomposition~\cite{Palhares:2010Droplets} which may lead to observable effects in colliding experiment. Although the real case encountered in future low energy collisions at RHIC and FAIR~\cite{Stephans:2006RHICLowEnergy,Senger:2004FAIR} depends on the detail of the phase evolution trajectory on the phase diagram, such as the time duration in the spinodal decomposition region~\cite{Arsene:2006Trajactory}, interface tension remains a key quantity in either of the two transition mechanisms. In astrophysics aspect, which is another important QCD test-ground, it is also discovered that a small value of interface tension in QCD matter favours the nucleation of quark matter during the early post-bounce stage of core collapse supernovae, allowing observable astrophysical phenomena~\cite{Palhares:2010Droplets,Palhares:2011AstrophysicsQCDBubbles}.

The interface effects are usually taken into account thermodynamically by interface tension and related quantities such as interface entropy and critical size of the bubble. Though restricted by sign problem at large chemical potential, lattice QCD  has tried to calculate the interface tension~\cite{deForcrand:2005Tension}. Besides, many effective model calculations have been performed to investigate interface effects~\cite{Heiselberg:1993Tension,Arsene:2006Trajactory,Randrup:2009PhaseTransition,Pinto:2012Tension,Peng:2010Tension,Mintz:2013Tension,Garcia:2013Tension,Palhares:2010Droplets,Lugones:2013Tension}, especially in the process of hadronization. The phenomenological  method of extracting the interface tension from the equation of states (EoS) of the system is proposed in Ref.~\cite{Randrup:2009PhaseTransition} by adding density gradient term to the energy density. The interface tension calculated in $2$-flavour and $2+1$ flavour NJL model and linear sigma model has been reported in Refs.~\cite{Pinto:2012Tension} and \cite{Palhares:2010Droplets,Arsene:2006Trajactory}, respectively. The latter also discuss the variation behavior of the critical bubble size with respect to quark chemical potential. These effective models give really low values of interface tension (from 5 to 30 $\textrm{ MeV/fm}{}^2$) at low temperature which favours nucleation process.
However, more recent $2+1$ flavour NJL model calculation~\cite{Lugones:2013Tension} gave a very large value ($145 \sim 165 \textrm{ MeV/fm}{}^2$) when considering the finite size effect. It is then imperative to revisit the interface tension problem with the NJL model.

Once the temperature dependence of interface tension is obtained, the associated entropy can be calculated straightforwardly by Maxwell relation. Even though the entropy of the quark and gluon system has been studied quite well (see, for example, Refs.~\cite{Costa:2008NJL,Blaizot:1999Entropy}), interface entropy has not yet been extensively discussed. Previous calculations showed that the entropy of the system does not increase with the decreasing of temperature and chemical potential. One usually regards the system to consist of either quarks and gluons, hadrons, or their mixture. According to the increasing entropy principle, the hadronization can not take place when lowering the temperature of the quark system. Then the entropy puzzle~\cite{Song:2010EntropyPuzzle} arises for the hadronization process.
In fact, interfaces emerge to form the quark droplets (the embryos of the hadrons). Such an interface production induces interface entropy which may contribute to that of the final hadron matter and strongly changes the entropy discontinuity scales and even its order of the two phases in the first order phase transitions. Therefore, considering the interface effect may shed light on clarifying the entropy puzzle~\cite{Song:2010EntropyPuzzle}.  It has been known that a quantity concerned with interface entropy production is the typical length scale characterizing the inhomogeneity of the system or how much interface is present in unit volume. Such a problem has been discussed with the thin-wall approximation~\cite{Randrup:2009PhaseTransition}. However, the thin-wall approximation may not be well justified at low temperatures and may overestimate the critical bubble size. Then, it is also beneficial to compare the thin-wall results with those from the improved method~\cite{Dunne:2005BeyondThinWall} and assess the influence of this change to the first order phase transition.

In the present work, we calculate the interface tension, the critical bubble size and the entropy density by 2+1 flavour Nambu-Jona-Lasino (NJL) model. It is found that the interface modification to the entropy density is important in mid-low temperature region. It shows that the thin-wall approximation overestimate the critical bubble size at low temperature. Furthermore, the value order of the entropies of the two phases in the hadronization process may be reversed so that the entropy puzzle can be solved if one improves the thin-wall approximation.

The paper is organized as follow. Following this introduction, we describe briefly the NJL model for the three flavour quark system in Section II, and give detailed description of the obtained phase diagram in Section III. In section IV and V we describe the interface thermodynamics and methods to extract the interface tension from EoS, respectively. The numerical results and discussion of the interface entropy and critical bubble radius are presented in Section VI. In Section VII, we assess the improvement on the thin-wall approximation and its influence. Finally, we summarize the present work and give some remarks in Section VIII.

\section{2+1 Flavour NJL Model}
The Lagrangian of three flavour quark NJL model at finite temperature and finite chemical potential is usually written as
\begin{widetext}
\begin{equation}
{\cal{L}} = \bar{\psi}(i\gamma^{\mu}\partial_{\mu}
         +\gamma^0\hat{\mu}-\hat{m})\psi
     +G\sum_{k=0}^{8}[(\bar{\psi}\lambda_k\psi)^2 +(\bar{\psi}i\gamma^5\lambda_k\psi)^2]
   - K\sum_{k=0}^{8}\{\det_f{\bar{\psi}(1+\gamma^5)\psi} +\det_f{\bar{\psi}(1-\gamma^5)\psi}\} \, ,
\end{equation}
\end{widetext}
where $\psi$ stands for the three-flavour quark field
\begin{equation}
\psi = {(\psi_u, \psi_d, \psi_s)}^{t} \, ,
\end{equation}
$\hat{m} = \textrm{diag}(m_u,m_d,m_s)$ is the current quark mass matrix in a diagonal form. The difference in mass between $u$ and $d$ quark is neglected here, so the isospin symmetry is preserved. The flavour $SU(3)$ symmetry is broken explicitly by a larger strange quark mass, i.e., $m_{s} \neq m_{u} = m_{d}$. The term $\gamma^0 \hat{\mu}$ serves as the lagrangian multiplets for the conservation of quark number of each flavour, taking into account their dependence on finite chemical potential. The chemical potentials for different quarks are set to be equal to each other for simplicity. $G$ and $K$ are the strength of the four-fermion interaction and the six-fermion interaction, respectively. The $\lambda_{k}^{}\, (k=1,2, \cdots,8 )$ in the four-fermion term are the eight Gell-Mann matrices and $\lambda_{0}^{} = \sqrt{\frac{2}{3}} \bf{I} $. The six-fermion term breaks the $U_{A}(1)$ symmetry but leaves $SU(3)_L \times SU(3)_R$ symmetry intact.

Taking the mean field approximation and treating the chiral condensate as a global classical field $\phi _{i}^{}= \langle \bar{\psi}_{i}^{} \psi_{i}^{} \rangle$ ($i = u, d, s$), the mean field lagrangian reads
\begin{eqnarray}
{\cal{L}}_{MF}^{} & = &
\bar{\psi}(i\gamma^{\mu}\partial_{\mu}+\gamma^0\hat{\mu}-\hat{M})\psi \nonumber \\
  & & - 2G({\phi_u}^2+{\phi_d}^2+{\phi_s}^2) + 4K\phi_u\phi_d\phi_s,
\end{eqnarray}
where $M_i= m_i-4G\phi_i+2K\phi_j \phi_k$ is the constituent quark mass. The grand potential can be obtained from the lagrangian following the standard procedure~\cite{Kapusta:2006Book},
\begin{eqnarray}
\Omega(T,\mu,\phi_i) & = & 2G({\phi_u}^2+{\phi_d}^2+{\phi_s}^2) -4K\phi_u\phi_d\phi_s  \nonumber  \\
& & - 2N_c\sum_{i=u,d,s}\int \frac{\textrm{d} p^3}{{(2\pi)}^3} \Big\{ \Theta(\Lambda^2-p^2)E_{i}  \nonumber \\
& & \;\; + T\ln{[1+\exp{(-\frac{E_i-\mu}{T})}]}  \nonumber \\
& & \;\; + T\ln{[1+\exp{(-\frac{E_i+\mu}{T})}] } \Big\} \, ,
\end{eqnarray}
where $E_i=\sqrt{p^2+{M_i}^2}$ is the single particle energy and the step function serves as a cut off to eliminate the ultraviolet  divergence.

The equilibrium condition requires
\begin{equation}
    \frac{\partial\Omega}{\partial\phi_u} = 0\, , \quad   \frac{\partial\Omega}{\partial\phi_d} = 0\, , \quad  \frac{\partial\Omega}{\partial\phi_s} = 0\, .
\end{equation}
By solving these equations, the chiral condensation can be determined self-consistently. Then, the grand potential is obtained by substituting the equilibrium values of condensations in the expression $\Omega(T,\mu,\phi_i)$. Because the difference between the $u$ and $d$ quark current masses is neglected, it follows that the $u$-quark and $d$-quark condensations are identical and the three equations reduce to two. These coupled nonlinear equations have more than one set of solutions and one can study the phase transition by analyzing the evolution of the solutions.

\section{Numerical Results of the Phase Diagram}

We have solved the coupled equations of the three flavour quark NJL model with the commonly adopted parameters ($m_u= m_d=5.5 \textrm{ MeV}, m_s=140.7 \textrm{ MeV}, \Lambda=602.3 \textrm{ MeV}, G = 1.835\Lambda^{-2}$ and $K = 12.36\Lambda^{-5}$ )~\cite{Rehberg:1996NJLParameters} and the homotopy method~\cite{Wang:2012Homotopy} to get all the solutions in the coexistence region.

\begin{figure}
\centering
\includegraphics[width=0.9\linewidth] {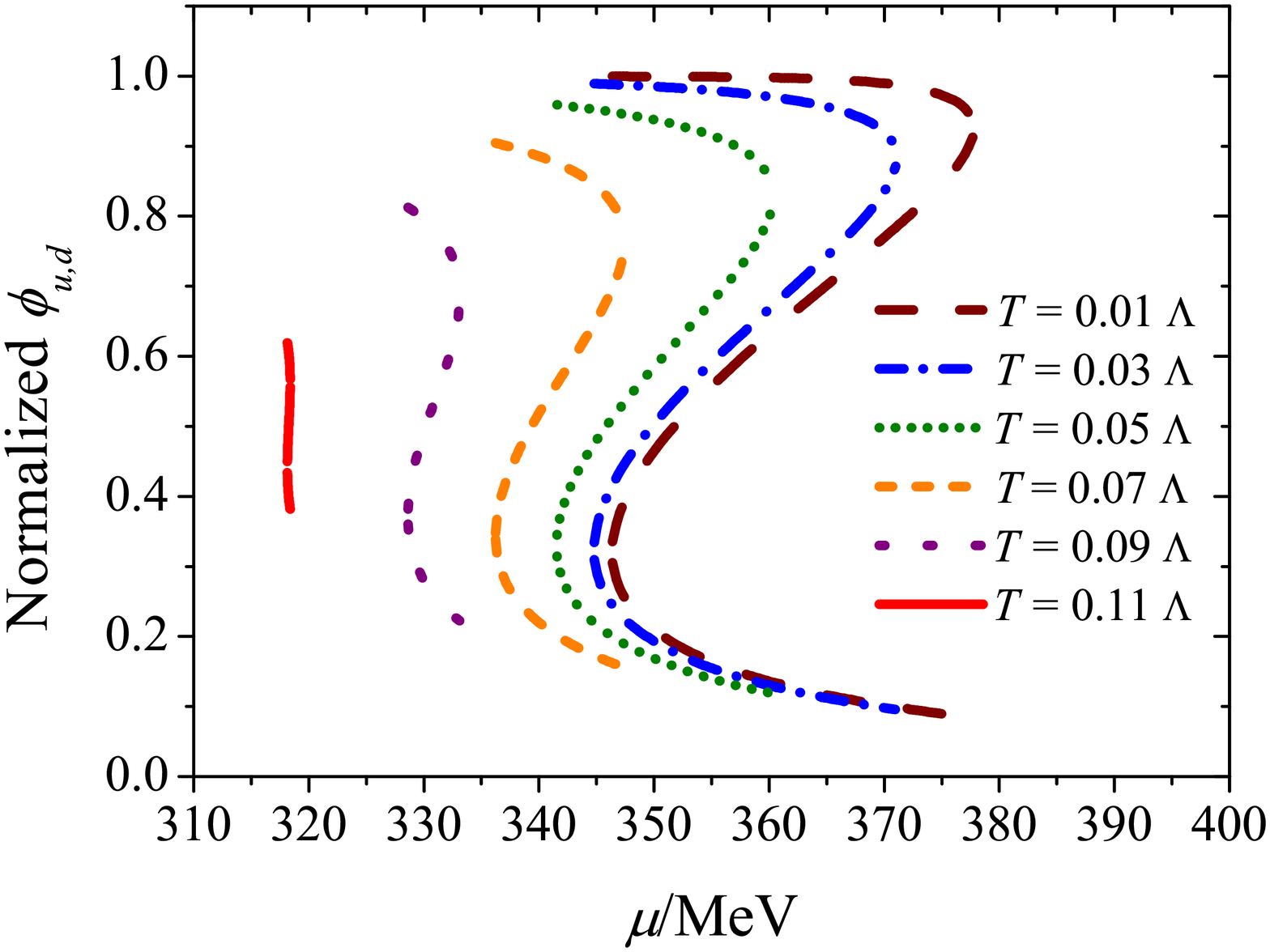}\\
\caption{(color online) Calculated variation behavior of the $u$ and $d$ quark condensation with respect to quark chemical potential at several temperatures (normalized by the condensation at zero temperature and zero chemical potential $\phi_{u,d}(T=0,\mu=0) = -{(242 \textrm{ MeV})}^3$ ).}\label{ud-cond}
\end{figure}

\begin{figure}
\centering
\includegraphics[width=0.9\linewidth]{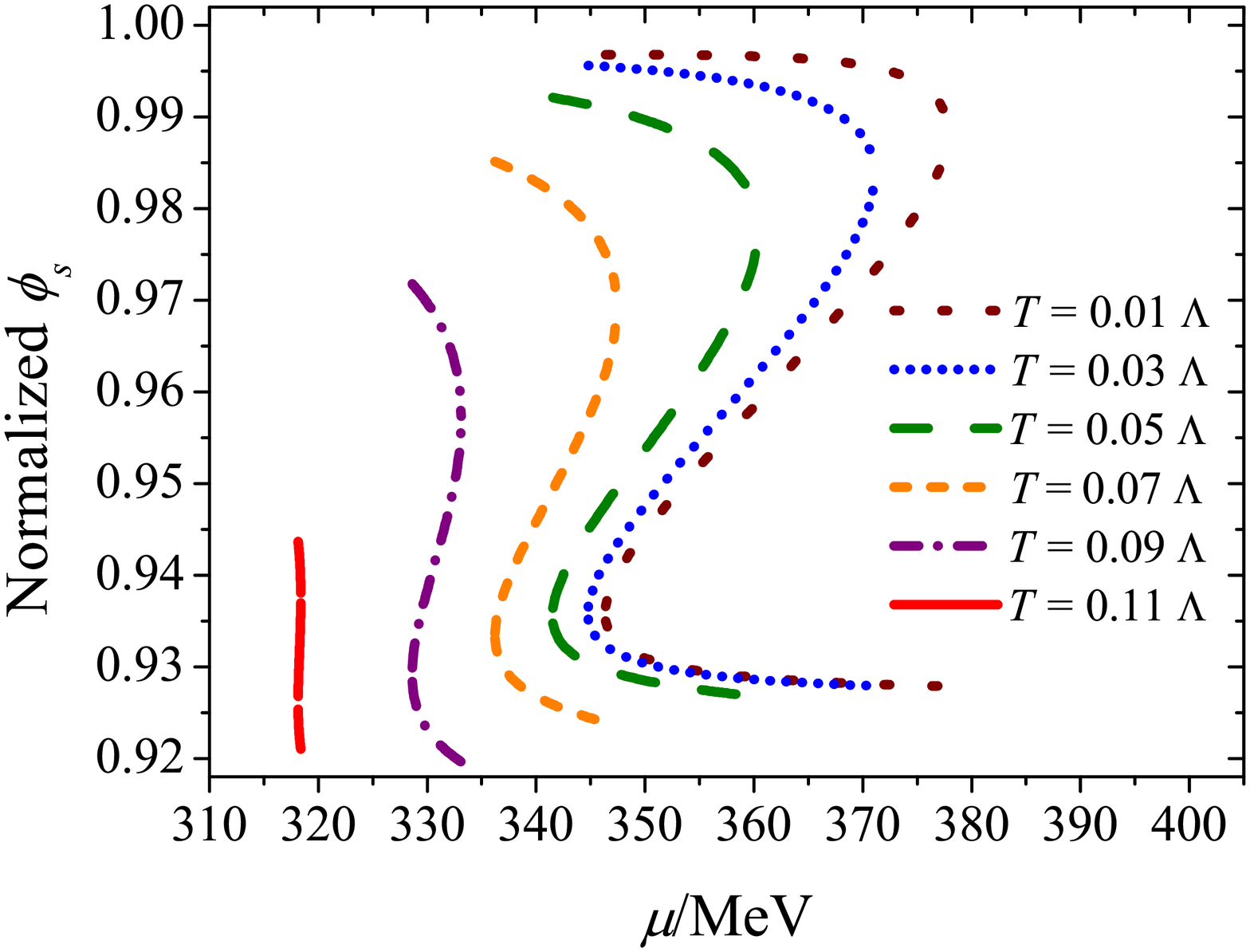}\\
\caption{(color online) Calculated variation behavior of the $s$ quark condensation with respect to quark chemical potential at several temperatures (normalized by the condensation at zero temperature and zero chemical potential $\phi_s(T=0,\mu=0) = -{(258 \textrm{ MeV})}^3$). }\label{s-cond}
\end{figure}

Fig.~\ref{ud-cond} and Fig.~\ref{s-cond} show the evolution behaviors of the chiral condensation with respect to chemical potential in the coexistence region at various temperatures (below the CEP temperature), corresponding to $u, d$ quark condensation and $s$ quark condensation, respectively. We see evidently from the figures that, at lower chemical potential, the Nambu phase (DCSB phase), is the only possible homogeneous state of the system. At higher chemical potential, only the dynamical chiral symmetry restored Wigner phase (DCS phase) is allowed. In the intermediate region where a chemical potential driven first order phase transition emerges, the condensations of all quarks are multi-valued. The upper branch and lower branch is the continuation of the Nambu phase and Wigner phase into the coexistence region while the intermediate line travelling backwards connecting the Nambu phase and Wigner phase is called the intermediate phase here. Moreover, with temperature going higher, the coexistence region moves to lower chemical potential with the condensations for all quarks dropping down and the backbendings of the curves getting weaker.

\begin{figure}
\centering
\includegraphics[width=0.9\linewidth]{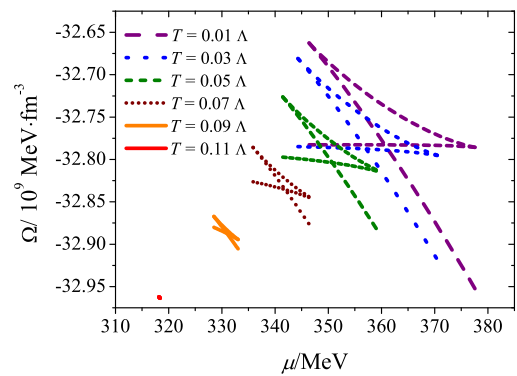}\\
\caption{ (color online) The grand potential in the coexistence region with respect to quark chemical potential at several  temperatures.}\label{Grand-Potential}
\end{figure}

The grand potentials corresponding to these two phases at various temperatures are illustrated in Fig.~\ref{Grand-Potential}. The figure shows apparently that the coexistence region (multi-values region) shrinks rapidly with the increasing of temperature and vanishes when the CEP temperature is reached. One can also notice that,  when the chemical potential is below a certain value, the Nambu (DCSB) phase is the stable one since it corresponds to the global minimum while the Wigner (DCS) phase is metastable; in case of above the certain chemical potential, the Wigner phase gets stable and the Nambu phase becomes metastable. It is clearly from Fig~\ref{Grand-Potential}, that the second order derivative of $\Omega$ with respect to chemical potential is negative for Nambu  and Wigner phases but positive for intermediate phase. As a result, Nambu and Wigner phases are stable against tiny disturbance in chemical potential while the intermediate phase is always unstable to such fluctuations. Although this argument precludes the existence of a homogeneous intermediate phase, this branch of solution is important in inhomogeneous system and can induce the appearance of the interface effect~\cite{Randrup:2009PhaseTransition} and, in turn, the phase transitions.

\begin{figure}
\centering
\includegraphics[width=0.9\linewidth]{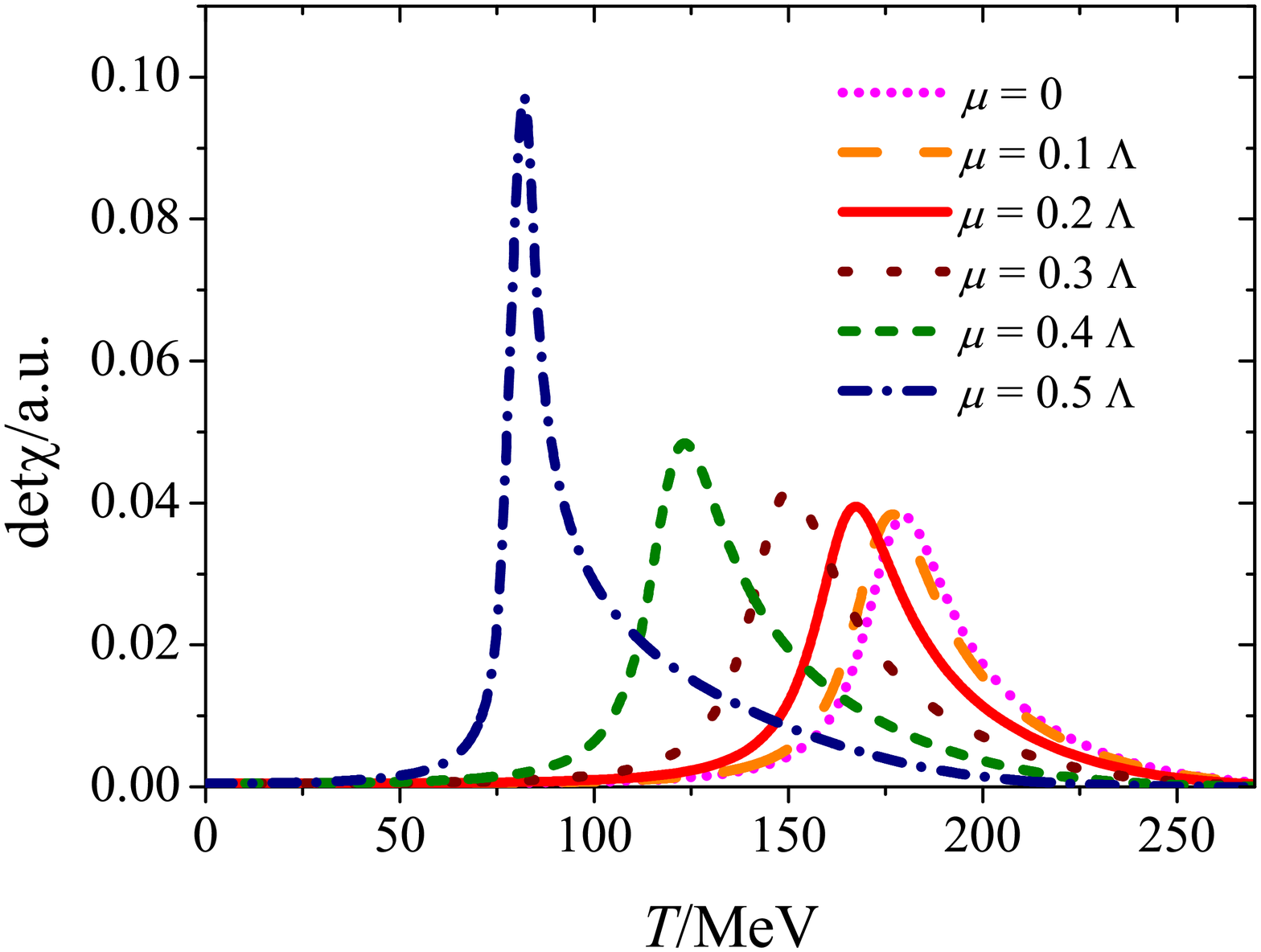}\\
\caption{(color online) Calculated variation behavior of the chiral susceptibility with respect to temperature at several quark chemical potentials.}\label{susept}
\end{figure}

With the above results, one can fix the boundary of the coexistence line in the $\mu$--$T$ plane. For completeness we have also calculated the crossover boundary by finding the temperatures that maximize the chiral susceptibility $\det(\chi_{ij})$ at each chemical potential, where $\chi_{ij}$ is defined by
\begin{equation}
\chi_{ij} = \frac{\partial\phi_i}{\partial m_j}.
\end{equation}
A detailed description of this chiral susceptibility can be found in Ref.~\cite{Jiang:2013NJL}. The obtained $\det(\chi_{ij})$ in terms of temperature at several chemical potentials ($\mu<\mu_{CEP}^{}$) is shown in Fig.~\ref{susept}. However, we will not discuss further the  property of the matter in the crossover region, as our main focus is the first order phase transition.

\begin{figure}
\centering
\includegraphics[width=0.9\linewidth]{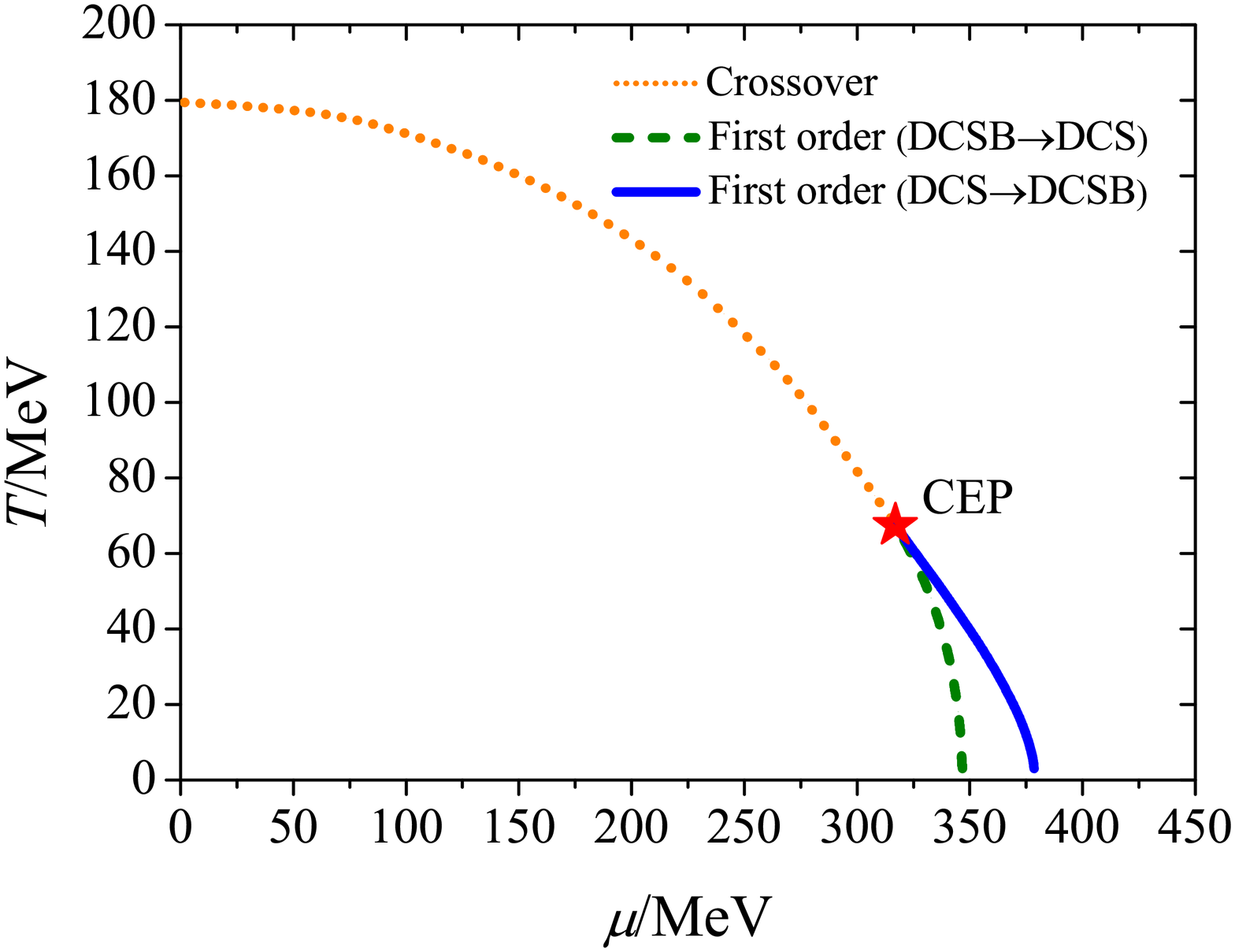}\\
\caption{(color online) Obtained phase diagram in terms of quark chemical potential and temperature. The CEP is labeled by a star.}\label{PD-mu-T}
\end{figure}

We can reach then the phase diagram in the $\mu$--$T$ plane. The obtained result is shown in Fig.~\ref{PD-mu-T}. The CEP determined on the $\mu$--$T$ plane is about $(316.3, 68.2)$~MeV with a quite large $\mu_{CEP}^{}/T_{CEP} ^{}\approx 4.6$, while this ratio obtained from the canonical lattice QCD and the  DSE approach is about $1$~\cite{Qin:2011DSE,Zong:2009DSE,deForcrand:2006Lattice,CEP:2002Lattice}. Such an inconsistency arises from the fact the the confinement length in the NJL model is zero~\cite{Qin:2011DSE}.

\begin{figure}
\centering
\includegraphics[width=0.9\linewidth]{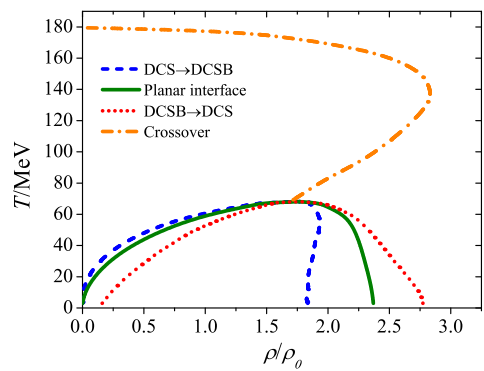}\\
\caption{(color online) Obtained phase diagram in terms of baryon density and temperature. $\rho_{0}^{} = 0.17\textrm{ fm}^{-3}$ is the saturation baryon density of nuclear matter.}\label{PD-rhoB-T}
\end{figure}

Considering the relation between the baryon density and the coexistence chemical potential
\begin{equation}
\rho_{B}^{} = \frac{1}{3} \, \rho_{q}^{} =  - \frac{1}{3} {\Big( \frac{\partial\Omega_{i}}{\partial\mu} \Big) }_{\mu=\mu_{co}},
\end{equation}
where ``$i$" stands for the ``Nambu" or ``Wigner" phase, we can translate the phase diagram on the $\mu$--$T$ plane into that on the $\rho_{B}^{}$--$T$ plane. The obtained result is diaplayed   in Fig.\ref{PD-rhoB-T}.
The dashed, solid, and dotted line in Fig.~\ref{PD-rhoB-T} stand for the results in three cases of geometry of phase transition: transition from DCS to DCSB phase, planar interface, and transition from DCSB to DCS phase, respectively. It is interesting that the chiral phase transition may take place on different coexistence lines when the system travels in different processes across the coexistence region (the reasons will be discussed in Section IV). When the system travels from DCSB phase to DCS phase, the phase transition takes place at  higher baryon density and with larger coexistence region; while it coexists with the opposite process in intermediate density region. A curved interface is generated in both of these cases. If the boundary of the two phases is set to be planar, it coincides with that given (traditionally) by the condition  $\Omega_{\textrm{Nambu}}(T,\mu) =  \Omega_{\textrm{Wigner}}(T,\mu)$.

\begin{figure}
\centering
\includegraphics[width=0.9\linewidth]{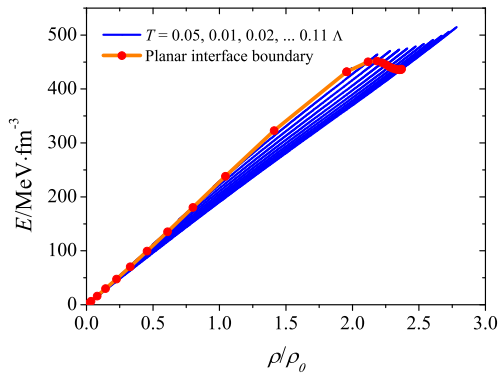}\\
\caption{(color online) Obtained phase diagram in terms of baryon density and energy density (the EoS). The blue lines are isothermal trajectories. $\rho_{0}^{} = 0.17\textrm{ fm}^{-3}$ is the saturation baryon density of nuclear matter.}\label{rho-E}
\end{figure}

The phase diagram can also be converted to the relation between the energy density and the baryon density (also referred to as the equation of state, EoS) in which the energy density reads
\begin{equation}
E(s,\rho) = Ts+\mu\rho+\Omega = -T\frac{\partial\Omega}{\partial T} -\mu\frac{\partial\Omega}{\partial\mu}+\Omega ,
\end{equation}
where the $s$ is the entropy density. The obtained result is shown in Fig.~\ref{rho-E}. The orange line connecting the red dot in the figure is the coexistence line in case of planar interface. As pointed out in Ref.~\cite{Arsene:2006Trajactory},  unlike the two previous cases, the $E$--$\rho$ relation is single valued, so it is easier to study the phase trajectory of the fireball on the $E$--$\rho$ plane. In the present work, the phase diagram in $E$--$\rho$ plane will be taken to set the parameters in interface tension calculations.

It should be noticed that the phase boundary we obtained now is too low in density compared with the lattice results~\cite{deForcrand:2006Lattice} and the conjectured phase diagram~\cite{Arsene:2006Trajactory}. In Ref.~\cite{Arsene:2006Trajactory} (the Fig. 1), the two phases coexist at nonzero densities when $T=0$ and the densities corresponding to the coexisting boundaries get lower as the temperature grows up and end up at a CEP with a small ${\mu_{\textrm{CEP}}^{}}/{T_{\textrm{CEP}}^{}}$ value. However,  present work gives that, at zero temperature, the DCS phase coexists with the DCSB  phase in density region $0$--$2.35\rho_{0}^{}$ and  the ratio ${\mu_{\textrm{CEP}}^{}}/{T_{\textrm{CEP}}^{}}$ takes a quite large value.  This is definitely not reasonable. Furthermore, the crossover boundary on the $T$--$\rho$ plane is multi-valued, which is also different from the conjectured phase diagram in Ref.~\cite{deForcrand:2006Lattice}. These two problems are also encountered in other work based on 2+1 flavour NJL model under different treatment~\cite{Sasaki:2008NJL,Costa:2008NJL}. We would attribute them to the absence of confinement effect in the model, but there may also be other aspects.

\section{Thermodynamics including the interface effect}

The above scheme and results capacite only for uniform bulk matters. In first order phase transition, spatial density inhomogeneity is in fact, significant  to understand properties of the mixed phase. The inhomogeneity leads to an interface tension that not only determines the critical bubble size, but also contributes extra entropy to the system during the nucleation process (in detail, the ``vaporization" or deconfinement when being heated and/or compressed, and the ``coagulation" or hadronization when being cooled and/or diluted). When the system is compressed continuously, interfaces finally collide with each other and the inhomogeneity disappears at the end of phase transition. Interface entropy and associated latent heat are released. However, in the nucleation process, which is also the focus of the present work, the system expands continuously. As a result, interfaces survive as the boundary of the finite size hadron and there is a net interface contribution to the entropy of the final state.

We begin our discussion with the thermodynamics of interface. The total derivative of the free energy of a bubble, incorporating the contribution of interface, is
\begin{equation}
\mathrm{d}F = -\Delta P\mathrm{d}V - S_{A}\mathrm{d}T + \gamma \mathrm{d}A \, ,
\end{equation}
with $\Delta P$ being the pressure difference of the two phases and $S_{A}$ being the interface entropy. $\gamma$, $A$ are the interface tension, interface area, respectively. Using the Maxwell relation and the extensity of entropy, we arrive at the interface entropy
\begin{equation}
S_{A} = A \cdot s_{A}^{} = A {\Big( \frac{\partial S_{A}}{\partial A } \Big)}_{V,T}^{} = - A {\Big( \frac{\partial \gamma}{\partial T} \Big)}_{V,A}^{} \, ,
\end{equation}
where $s_{A}^{}$ being interface entropy density, is the opposite of the partial derivative of $\gamma$ with respect to $T$.

The total entropy density of the system can be gained by summarizing the body entropy density and the interface entropy density. A straightforward result is
\begin{equation}
s_{tot}^{}(T) = s_{V}^{} + \frac{A}{V} s_{A}^{},
\end{equation}
where $V$ and $A$ are the volume of the system and the total area of interfaces within the system. The problem is then that a reasonable determination of the ratio $A/V$ depends on the dynamics of the interface generation, which is too complicated to be addressed completely and fundamentally in the present work. However, it is intuitively to assume that the total area of the interfaces, $A$, depends on how many bubbles are generated in the system after the phase transition. Assuming that the system is divided by lattice in which spherical bubbles are formed, we have then
\begin{equation}
A \approx \frac{\pi V}{2 \overline{R}} \, ,
\end{equation}
where $\overline{R}$ is the average of the radius of the bubbles. If the system enters the spinodal region, another possible way to overcome this difficulty is to treat $A/V$  as the typical length scale of the spinodal decomposition. It is pointed out in Ref.~\cite{Randrup:2009PhaseTransition} that the gradient term modified dispersion relation for the density fluctuation is of the form
\begin{equation}
\omega^2 \approx \alpha(T,\mu) k^2 + \beta  k^4 \, .
\end{equation}
In the spinodal decomposition region, $\alpha(T,\mu)$ is negative and the infinitesimal density fluctuation grows exponentially with time. Due to the presence of the $\beta k^4$ term, the growth of fluctuation of a typical wave number exceeds those of other modes and dominates the decomposition process. We can take the inverse of the wave length of this mode as an evaluation of $A/V$. However, this calculation is out of the scope of the present work which only deals with the nucleation region. Here we brutally assume that all the bubbles are generated uniformly with radius $R$, so the total entropy density is
\begin{equation} \label{Total-EntropyDensity}
s_{tot}^{}(T)= s_{V}^{}+ \frac{\pi}{2R} s_{A}^{}  \, .
\end{equation}
The spatial scale $R$ is characterized by the critical nucleation radius $R_{c}^{}$ that can be extracted by setting $\mathrm{d}T=0$ and $\mathrm{d}F=0$, i.e.,
$$\frac{\mathrm{d}V}{\mathrm{d}A} = \frac{\gamma}{\Delta P} \, , $$
We have then
\begin{equation} \label{Critical-R}
R_{c}^{} = \frac{2\gamma(T)}{-\Omega_{1}^{}(T,\mu)+\Omega_{2}^{}(T,\mu)  } \, .
\end{equation}
Because the fluctuation formation of smaller bubble is easier, we requires: when the system travels from the DCSB phase (hadron phase) to the DCS phase (quark and gluon phase), the two phases coexist at the highest chemical potential for the coexistence line on the $\Omega$--$\mu$ plane, while when the system expands and travels back across the coexistence region, the two phases coexist at the lowest possible chemical potential.

\section{Thin-Wall Approximation}

A phenomenological treat to the inhomogeneity is adding a gradient term in the free energy density~\cite{Randrup:2009PhaseTransition},
\begin{equation}
f({\bf r}) = \rho\mu+\frac{1}{2}C{(\nabla\rho)}^2.
\end{equation}
$C= \frac{a^2}{{\rho_{g}^{}}^2} E_{g}^{}$ is a constant determined by the typical density $\rho_{g}^{}$ and energy density $E_{g}^{}$ of the coexistence region and a measure of the thickness $a$ of the interface. In the present work, we take the same $a$ as that in Ref.~\cite{Garcia:2013Tension}, i.e., $a =0.33$~fm.  $\rho_{g}^{} = 1.83 \textrm{ fm}^{-3}$  (or quark density $5.5 \textrm{ fm}^{-3}$) is the baryon density at the CEP and $E_{g}^{} = 370 \textrm{ MeV} \textrm{fm}^{-3}$ for the energy density is chosen in the medial region at $\rho_{g}^{}$ in Fig.~\ref{rho-E}.

Equilibration condition requires a stationary free energy with respect to the variation of density distribution under the restriction of normalization
\begin{eqnarray}
 0 & = & \delta\int\{\rho({
\bf r})\mu_{T_0}^{}[\rho] + \frac{1}{2}C{(\nabla\rho)}^2 - \mu_{0}^{} \rho({\bf r})\}\mathrm{d}{\bf r} \, , \quad \\
 N & = & \int{\rho({\bf r})\mathrm{d}{\bf r}} \, .
\end{eqnarray}
Then, at the phase transition temperature $T_{0}^{}$, the equation of motion reads
\begin{equation}
\mu_{T_{0}^{}}[\rho] - \mu_{0}^{} - C\nabla^2\rho=0.
\end{equation}
In the planar interface case,
\begin{equation}
C{\Big( \frac{\mathrm{d}\rho}{\mathrm{d}x} \Big) }^2 - \Delta f_{T_{0}^{}} = 0
\end{equation}
is conserved with respect to $x$, in which  $\Delta f_{T_{0}^{}} = f_{T_{0}^{}}(\rho) - f_{M}^{}(\rho)$ is the difference between the uniform free energy density and the Maxwell construction of free energy, which is
\begin{equation}
f_{M}^{}(\rho)= f_{T_0}^{}(\rho_{L}^{})+\frac{f_{T_0}^{}(\rho_{H}^{})
-f_{T_0}^{}(\rho_{L}^{})}{\rho_{H}^{} - \rho_{L}^{}} (\rho-\rho_L),
\end{equation}
where $\rho_{H}^{}$ and $\rho_{L}^{}$ are the density for the two phases evaluated at the coexistence chemical potential. The Maxwell construction is chosen so as to fulfill the boundary condition that far away from the interface, the density should be uniform and approaches $\rho_{H}^{}$ and $\rho_{L}^{}$ on either side as if no interface is present. The interface tension connects to the free energy deficit per unit area on the interface, as usual,
\begin{eqnarray}  \label{InterfaceTension-T}
\gamma(T) & = & \int_{-\infty}^{\infty}{\Delta f_{T}^{} \mathrm{d}x} = \int_{\rho_{L}^{}}^{\rho_{H}^{}}{C{\Big( \frac{\mathrm{d}\rho}{\mathrm{d}x} \Big) }^2 \frac{\mathrm{d}x}{\mathrm{d}\rho}\mathrm{d}\rho}    \nonumber \\  & = & \int_{\rho_{L}^{}}^{\rho_{H}^{}}{\sqrt{\frac{\delta f_{T}^{}(\rho)}{C}} \mathrm{d}\rho}.
\end{eqnarray}
Note that the result above is the tension for infinite planar interface, but bubble interface formed in medium has a finite curvature. The above calculation can be generalized to the spherical case as
\begin{equation}
0 = \delta\int \Big\{ \rho(
r)\mu_{T_0}[\rho] + \frac{1}{2}C{(\frac{\mathrm{d}\rho}{\mathrm{d}r})}^2
 - \mu_0\rho(r) \Big\} r^2 \mathrm{d}r.
\end{equation}
The ``equation of motion" can be modified with a ``damping term" (this damping term is not important in the limit of thin-wall approximation $r\rightarrow\infty$) with
\begin{equation}
\frac{\mathrm{d}^2\rho}{\mathrm{d}r^2}+\frac{2}{r}\frac{\mathrm{d}\rho}{\mathrm{d}r} = -\frac{1}{C}\frac{\partial(-\Delta f)}{\partial\rho}.
\end{equation}
The conserved integral is gone and now we have to solve the density profile under the boundary condition in the first place. The calculation is then an improvement on the original thin-wall calculation. However, out of the regime of thin-wall approximation, determination of the radius and even the existence of the interface are ambiguous. These will be discussed in Section VII.

\section{Numerical Results of Interface Effects}

\begin{figure}
\centering
\includegraphics[width=0.9\linewidth]{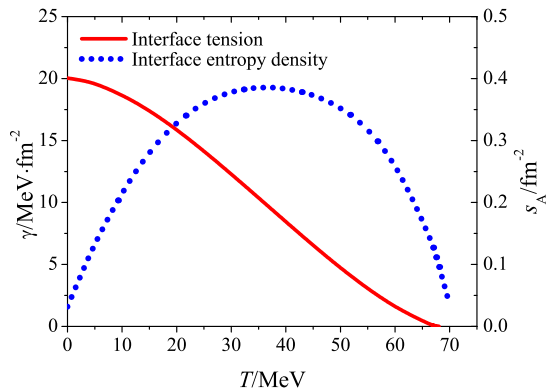}\\
\caption{(color online) Calculated variation behaviors of the interface tension and interface entropy density with respect to temperature.}\label{tension}
\end{figure}

\begin{figure}
\centering
\includegraphics[width=0.9\linewidth]{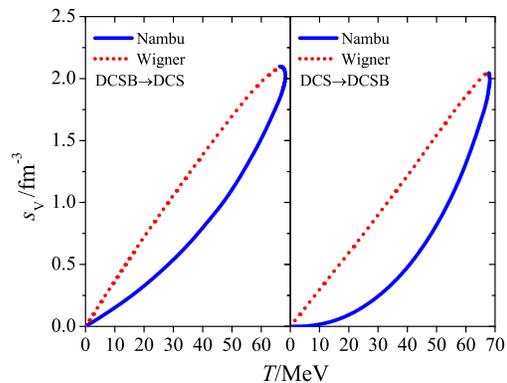}\\
\caption{(color online) Calculated body entropy density in terms of temperature. Left panel: DCSB $\rightarrow$ DCS (deconfinement). Right panel: DCS $\rightarrow$ DCSB (hadronization).}\label{body-entropy}
\end{figure}

The numerical results for interface tension and entropy are shown in Fig.~\ref{tension}. The red line in the figure displays the obtained variation behavior of the interface tension with respect to the temperature. The result can be fitted by a polynomial. The negative of the derivative of the fitted curve with respect to temperature gives the interface entropy density (shown as blue dotted line in the figure). It is apparent that the interface tension is about $20\;\textrm{MeVfm}^{-2}$ at $T=0$ and decreases monotonically with increasing temperature and approaches zero near the CEP temperature. Because the two phases practically make no difference with each other at the CEP, it is natural that the interface tension vanishes. The interface entropy density drops to zero at zero temperature as well as at the CEP temperature and have a broad peak at the intermediate temperature region $25 \sim 50$~MeV.
It is well known that the body entropy density of the two phases can be given as
\begin{equation}
s_{V,i}^{}(T)=-{\Big( \frac{\partial\Omega}{\partial T} \Big) }_{\mu = \mu_{co}} \, ,
\end{equation}
where ``$i$" denotes the ``Nambu" or ``Wigner" phase. The obtained result is showed in Fig.~\ref{body-entropy}. The left panel and right panel correspond to the processes from Nambu phase (DCSB) to Wigner phase (DCS) and from Wigner to Nambu phase respectively.

\begin{figure}
\centering
\includegraphics[width=0.9\linewidth]{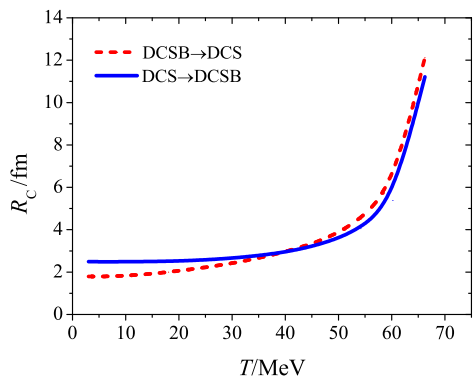}\\
\caption{(color online) Calculated critical bubble size in the two phase transition processes with respect to temperature.}\label{bubble-size}
\end{figure}

Fig.~\ref{body-entropy} shows evidently that there exists a  discontinuity in the entropy densities of the two phases during the first order phase transition, where the entropy density of the DCS phase is always higher than that of the DCSB phase in both heating and hadronization process. Such a discontinuity behavior in the DCS to DCSB transition is definitely not reasonable. Because if the nucleation process exists, the presence of a droplet in unit volume should increase the total entropy so that this process is favoured by the maximum entropy requirement. We expect then the entropy density of the DCSB phase becomes greater somewhere than that of the DCS phase in the hadronization process. To solve this problem, we take into account the interface generation and the associated interface entropy. To include the interface effect into the total entropy density of the system, we need the spatial scale of the bubbles. Basing on the discussion in last section, with Eq.~(\ref{Critical-R}) and the obtained results of the related quantities, we can get the critical bubble size for each phase transition process. The obtained results are shown in Fig.~\ref{bubble-size}. It is obvious that the bubble size is small and increases slowly at low temperature. In the vicinity of the CEP temperature the bubble size grows abruptly and gets divergent at the CEP temperature which represents the accomplishment of first order phase transition region. This result indicates that the contribution of the interface to the entropy is quite prominent in low and medium temperatures.

\begin{figure}
\centering
\includegraphics[width=0.9\linewidth]{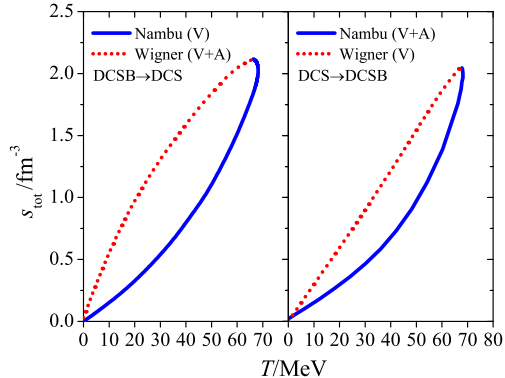}\\
\caption{(color online) Calculated total entropy density in terms of temperature. Left panel: DCSB $\rightarrow$ DCS. Right panel: DCS $\rightarrow$ DCSB (hadronization). The ``V" and ``A" in the brackets stand for the body and interface contribution respectively.}\label{Total-entropy-thin-wall}
\end{figure}

Fig.~\ref{Total-entropy-thin-wall} illustrates the obtained total entropy density of the coexistence phases in terms of the temperature. The left panel corresponds to the process from DCSB to DCS phase, where the initial phase is the uniform DCSB phase without interface contribution and the final phase is the one consisting of droplet  DCS phase. The interface contribution enhances the entropy density discontinuity and this effect is more prominent in the middle temperature, where the interface entropy density is large and the bubble size is small. The right panel shows the result in the process from uniform DCS phase to DCSB phase composed of hadrons (quark droplets), which mimics   the process of hadronization from quark gluon matter. Comparing Fig.~\ref{Total-entropy-thin-wall} with Fig.~\ref{body-entropy}, one can notice that the interface contribution reduces the entropy density discontinuity in the hadronization process, but the problem of the value order still exists.

\section{Improvement on the Thin-wall Approximation}

It has been mentioned in section V that the thin-wall approximation may break down when the thickness of the interface is of the same order as the radius of the bubble, which is quite likely at low temperature. To investigate the effect of its possible breaking down, we solve the equations of motion with damping term in section V under proper boundary conditions and obtain the radial density profile of the bubble at various temperatures. The critical bubble size are determined by the radius corresponding to the steepest change of the density profile.

\begin{figure}
\centering
\includegraphics[width=0.9\linewidth]{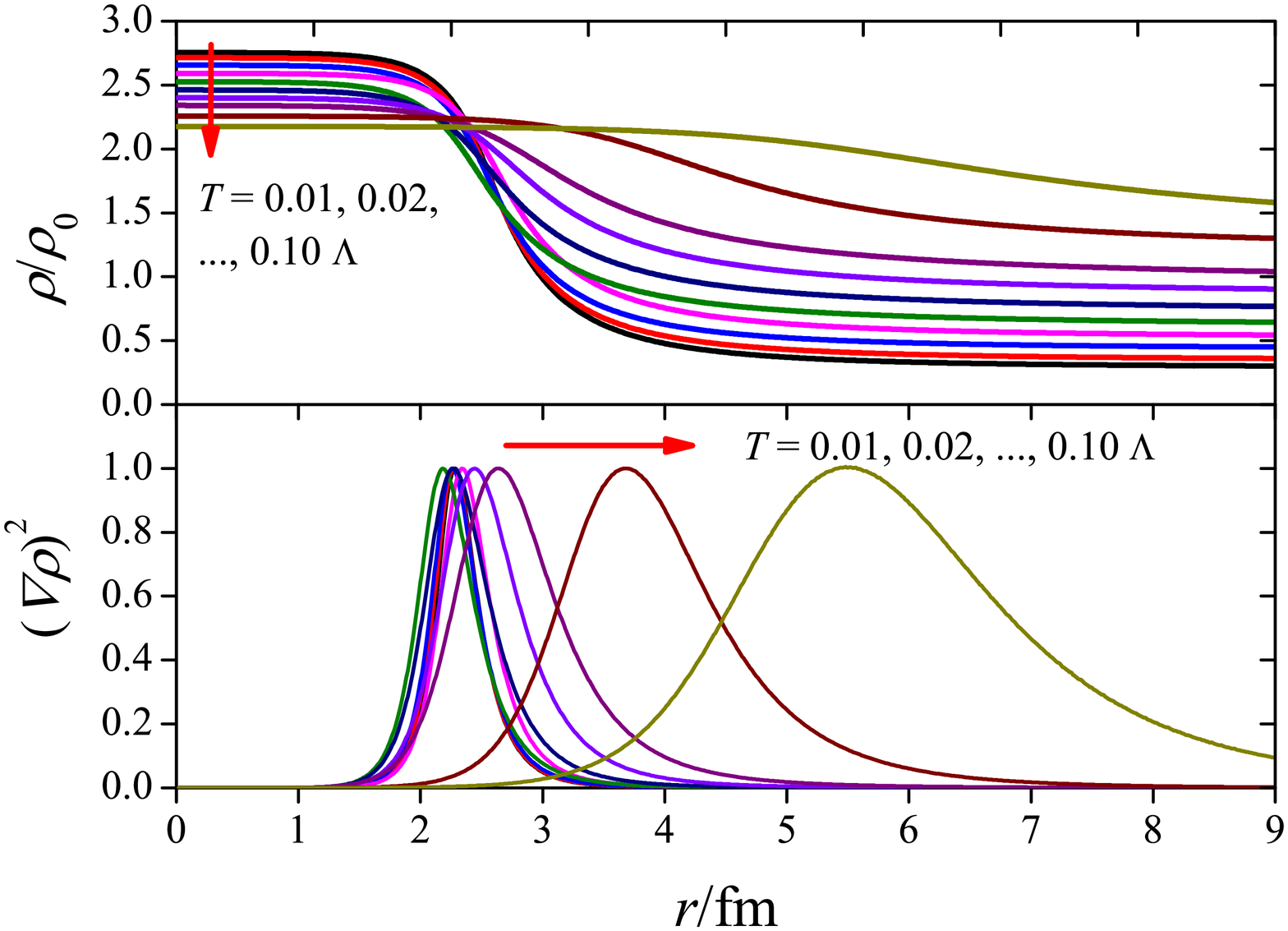}\\
 \caption{(color online) Calculated bubble density profile (upper panel) and the normalized square of the density gradient (lower panel) at several temperatures in the transition from Nambu phase to Wigner phase.}\label{Density-Profile-NtoW}
\end{figure}

\begin{figure}
\centering
\includegraphics[width=0.9\linewidth]{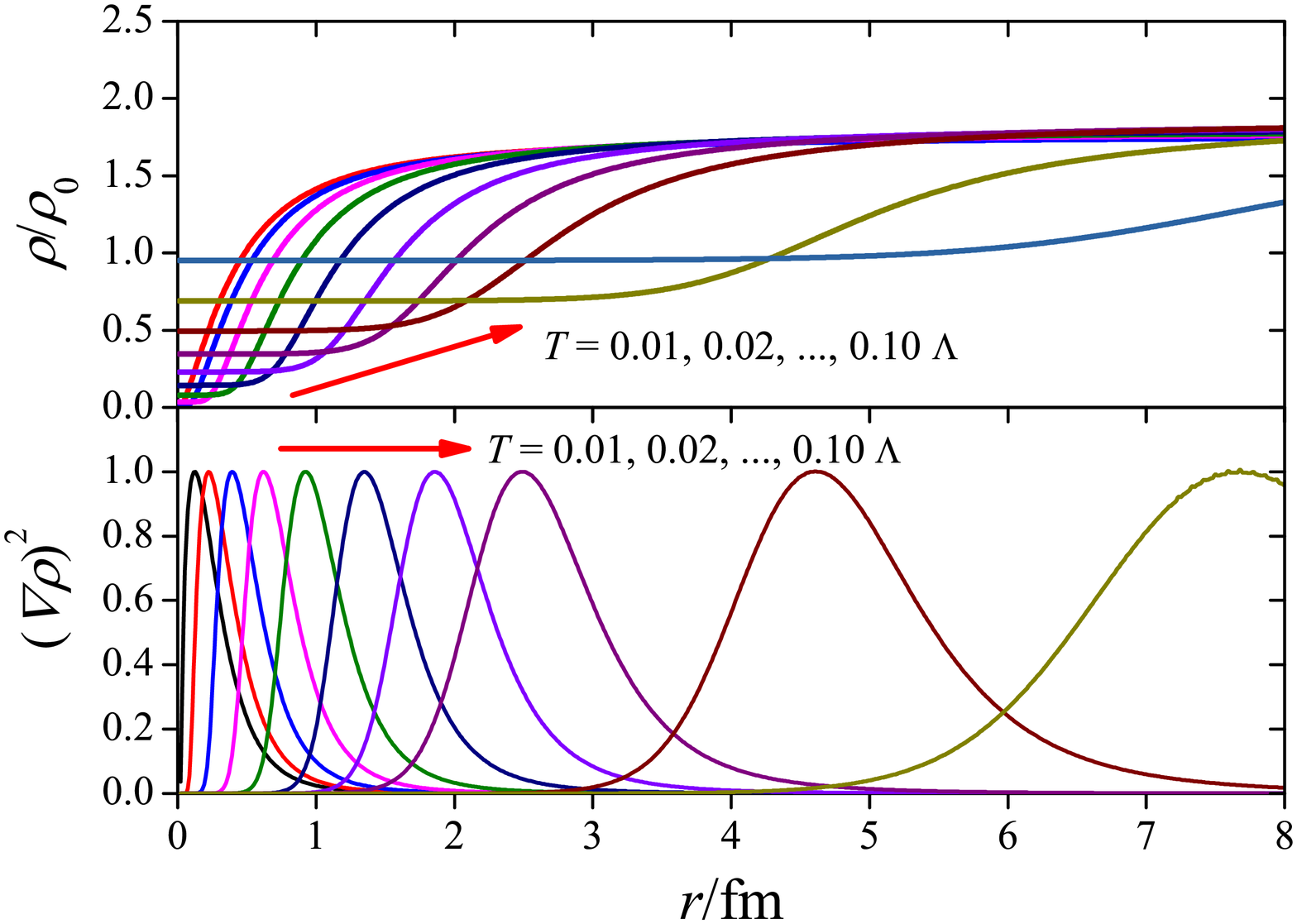}\\
\caption{(color online) Calculated bubble density profile (upper panel) and the normalized square of the density gradient (lower panel) at several temperatures in the transition from Wigner phase to Nambu phase.}\label{Density-Profile-WtoN}
\end{figure}

\begin{figure}
\centering
\includegraphics[width=0.9\linewidth]{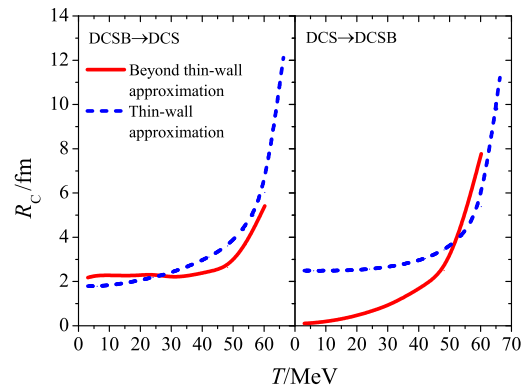}\\
\caption{(color online) Calculated critical bubble sizes in the two processes with the improved thin-wall approximation (red lines) and the comparison with the results within the original thin-wall approximation (blue lines).}\label{beyond}
\end{figure}

\begin{figure}
\centering
\includegraphics[width=0.9\linewidth]{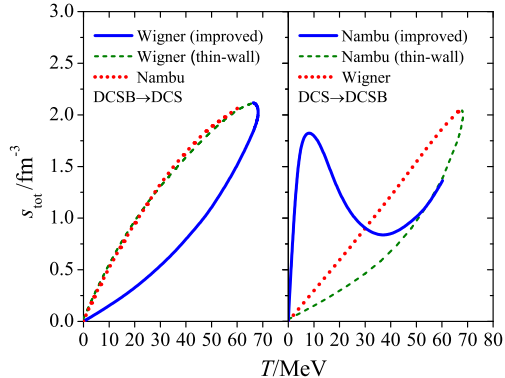}\\
 \caption{(color online) Total entropy density calculated with the improved thin-wall results of $R_c$. Left panel: DCSB $\rightarrow$ DCS (deconfinement). Right panel: DCS $\rightarrow$ DCSB (hadronization). For comparison, the original thin wall results are also displayed for each process  (thinner dotted lines). }\label{beyond-entropy}
\end{figure}

In Fig.~\ref{Density-Profile-NtoW} and Fig.~\ref{Density-Profile-WtoN}, we present the calculated density profile (upper panel for each figure) and the normalized square of the density gradient (lower panel for each figure) during the nucleation process from Nambu phase to Wigner phase and the opposite, respectively. We can further read off the critical bubble size from the lower panel directly. The obtained $R_c$ are displayed as the red lines in Fig.~\ref{beyond}, in which we also display the previous results (blue dashed lines) for comparison. It shows evidently that, during the heating process (from the DCSB to DCS phase), the original thin-wall results are in agreement with that in the improved calculation. However, in the transition from DCS to DCSB phase, the critical bubble size at low temperature calculated in the improved calcualtion diminishes quickly with decreasing temperature, while the original thin-wall results are not strongly affected by temperature.

This brings a major difference in the variation behaviors of the total entropy densities. If we assume that the interface tension does not rely strongly on the thin-wall approximation and replace the thin-wall results of $R_c$ in Eq. (\ref{Total-EntropyDensity}) by the ones  in the improved thin-wall approximation (as shown in Fig.~\ref{beyond}), the modified total entropy densities can be calculated straightforwardly with its results shown in Fig.~\ref{beyond-entropy}. As expected, the entropy of heating process is not strongly affected. However, we are surprised to find that, when the improved results of $R_c$ are plugged in, the entropy of Nambu phase in the hadronization process generates a lump at low temperature and the value order of the total entropy densities of Nambu and Wignre phases reverses, which makes the hadronization process practical in increasing entropy principle. The present results shed light on the entropy puzzle \cite{Song:2010EntropyPuzzle} by incorporating interface entropy and improving the thin-wall approximation.

\section{Summary}
In summary, we have investigated the QCD phase diagram in NJL model and pay special attention to the interface effects in the first order phase transition. Making use of the thin-wall approximation, the temperature dependence of the interface tension is obtained and the interface entropy density and critical bubble size are extracted. We show that the interface contribution to the total entropy is significant in the mid-low temperature region and changes the entropy density discontinuity scale between the initial and finial phase. We also discover that the thin-wall approximation drastically overestimates the critical bubble size at low temperature in the hadronization process. By rendering the entropy density of the  hadron phase obtained by improving the thin-wall approximation, we get that the total entropy density of the DCSB (confinement) phase exceeds that of the DCS (deconfinement) phase. The entropy puzzle in the hadronization process is then solved.

It should be remarked that a more realistic description of QCD phase transition should incorporate confinement. However, NJL model does not contain any information about confinement. This may be the reason of too low a density obtained for the coexistence of DCSB-hadron and DCS-quark phases, compared to the conjectured and calculated values in Refs.~\cite{Arsene:2006Trajactory} and \cite{deForcrand:2006Lattice}. We are then carrying out similar calculations in the framework of the Polyakov loop improved NJL  (PNJL) model~\cite{Fukushima:2008PNJL,Fu:2008PNJL} model, which takes the confinement effect statistically into account.
We are also going beyond the thin-wall approximation to study the spinodal process of the first order phase transition.

\bigskip


The work was supported by the National Natural Science Foundation of China under Contract Nos.\ 10935001, 11075052 and 11175004, the National Key Basic Research Program of China under Contract No.\ G2013CB834400.
%


\end{document}